\documentclass[conference]{IEEEtran}
\usepackage{caption}
\IEEEoverridecommandlockouts
\usepackage{cite}
\usepackage{amsmath,amssymb,amsfonts}
\usepackage{algorithm}      
\usepackage{algorithmic}

\usepackage{graphicx}
\usepackage{textcomp}
\usepackage{xcolor}
\usepackage{float}
\usepackage{bm}
\usepackage{soul}
\usepackage{subcaption}
\usepackage{stfloats}
\usepackage{siunitx}
\def\BibTeX{{\rm B\kern-.05em{\sc i\kern-.025em b}\kern-.08em
    T\kern-.1667em\lower.7ex\hbox{E}\kern-.125emX}}
\begin{document}

\title{IDSOR: Intensity- and Distance-Aware Statistical Outlier Removal for Weather-Robust LiDAR Point Clouds}

\author{
\IEEEauthorblockN{
    Chenyang Yan\IEEEauthorrefmark{1}
    Mats Bengtsson \IEEEauthorrefmark{1}
    } 
\IEEEauthorblockA{  
\IEEEauthorrefmark{1}KTH Royal Institute of Technology
}
\thanks{Funded by the European Union through Trafikverket. Views and opinion expressed are however those of the author(s) only and do not necessarily reflect those of the European Union or the Europe’s Rail Joint Undertaking. Neither the European Union nor the granting authority can be held responsible for them. The project FA6 FutuRe is supported by the Europe’s Rail Joint Undertaking and its members.}

}
\maketitle

\begin{abstract}
LiDAR point clouds captured in rain or snow are often corrupted by weather-induced returns, which can degrade perception and safety-critical scene understanding. This paper proposes Intensity- and Distance-Aware Statistical Outlier Removal (IDSOR), a range-adaptive filtering method that jointly exploits intensity cues and neighborhood sparsity. By incorporating an empirical, range-dependent distribution of weather returns into the threshold design, IDSOR suppresses weather-induced points while preserving fine structural details without cumbersome manual parameter tuning. We also propose a variant that uses a previously proposed method to estimate the weather return distribution from data, and integrates it into IDSOR. Experiments on simulation-augmented level-crossing measurements and on the Winter Adverse Driving dataset (WADS) demonstrate that IDSOR achieves a favorable precision--recall trade-off, maintaining both precision and recall above 90\% on WADS.
\end{abstract}


\section{Introduction}
\label{sec:1}
Light Detection and Ranging (LiDAR) is a sensing technology that employs pulsed laser signals to perform tasks such as distance measurement, three-dimensional point cloud generation, environmental modeling, and direction-of-arrival estimation. Owing to its high spatial resolution and accuracy, LiDAR has been widely adopted in a variety of applications, including autonomous driving, surveying and terrain mapping, and obstacle detection across diverse operational scenarios~\cite{li2020lidar,mulder2011use,gargoum2017automated,yue2024lidar,yan2026obstacledetectionlevelcrossings}.

Under adverse weather conditions, such as rain and snow, LiDAR point clouds can be contaminated by additional returns originating from rain and snow particles, which are commonly referred to as outliers in point cloud and image processing. These outliers typically appear in the vicinity of the LiDAR sensor and become denser at shorter ranges. The number of such reflections decreases with increasing distance from the sensor, with a maximum detectable range of approximately 10--20~m depending on the characteristics of the weather particles \cite{charron2018noising}.

To suppress weather-induced points in LiDAR data, several geometric outlier-removal schemes have been investigated, including Distance Outlier Removal (DOR), Statistical Outlier Removal (SOR), Dynamic Radius Outlier Removal (DROR), and Distance-Statistical Outlier Removal (DSOR)~\cite{rusu20113d,balta2018fast,charron2018noising,kurup2021dsor}. DOR typically labels a point as an outlier when its local neighborhood is sparse, for example, if the number of neighbors within a fixed radius falls below a preset threshold, indicating an isolated spatial configuration~\cite{rusu20113d,balta2018fast}. In contrast, SOR evaluates the mean distance from each point to its $k$ nearest neighbors (KNN) and removes points whose mean distance deviates significantly from the global statistics, commonly using a standard-deviation-based criterion~\cite{rusu20113d}. DROR extends the fixed-radius strategy by dynamically adjusting the search radius as a function of range, with the goal of maintaining a comparable neighborhood size across distances while discarding points that remain insufficiently supported~\cite{charron2018noising}. Building upon these ideas, DSOR incorporates the range-dependent sampling density of LiDAR by adapting the statistical decision threshold with respect to distance, thereby mitigating the over-removal of far-range points~\cite{kurup2021dsor}. Between DSOR and DROR, the results reported in~\cite{kurup2021dsor} indicate that DSOR typically achieves higher recall but lower precision than DROR, reflecting a stronger ability to suppress outliers at the expense of inadvertently removing true target points.

Despite their effectiveness and simplicity, existing geometric outlier-removal methods operate solely on the spatial distribution of points, relying on coordinates and neighborhood geometry, while ignoring the intensity information available in LiDAR measurements. Such intensity cues can be informative for distinguishing weather-induced returns from true scene surfaces. An extension of DSOR, termed Dynamic Distance–Intensity Outlier Removal (DDIOR), incorporates intensity information by defining a distance- and intensity-dependent threshold based on the “near-is-dense, far-is-sparse” characteristics of LiDAR point clouds and the statistical properties of snow-induced noise~\cite{wang2022scalable}. However, DDIOR requires manual tuning of weighting parameters at each distance interval (e.g., every 10 m as adopted in the original study) to balance the contributions of distance and intensity in the threshold design, which can be cumbersome in practice.

\begin{figure*}[!t]
\centering
\subfloat[Rainy]{\includegraphics[width=0.32\textwidth]{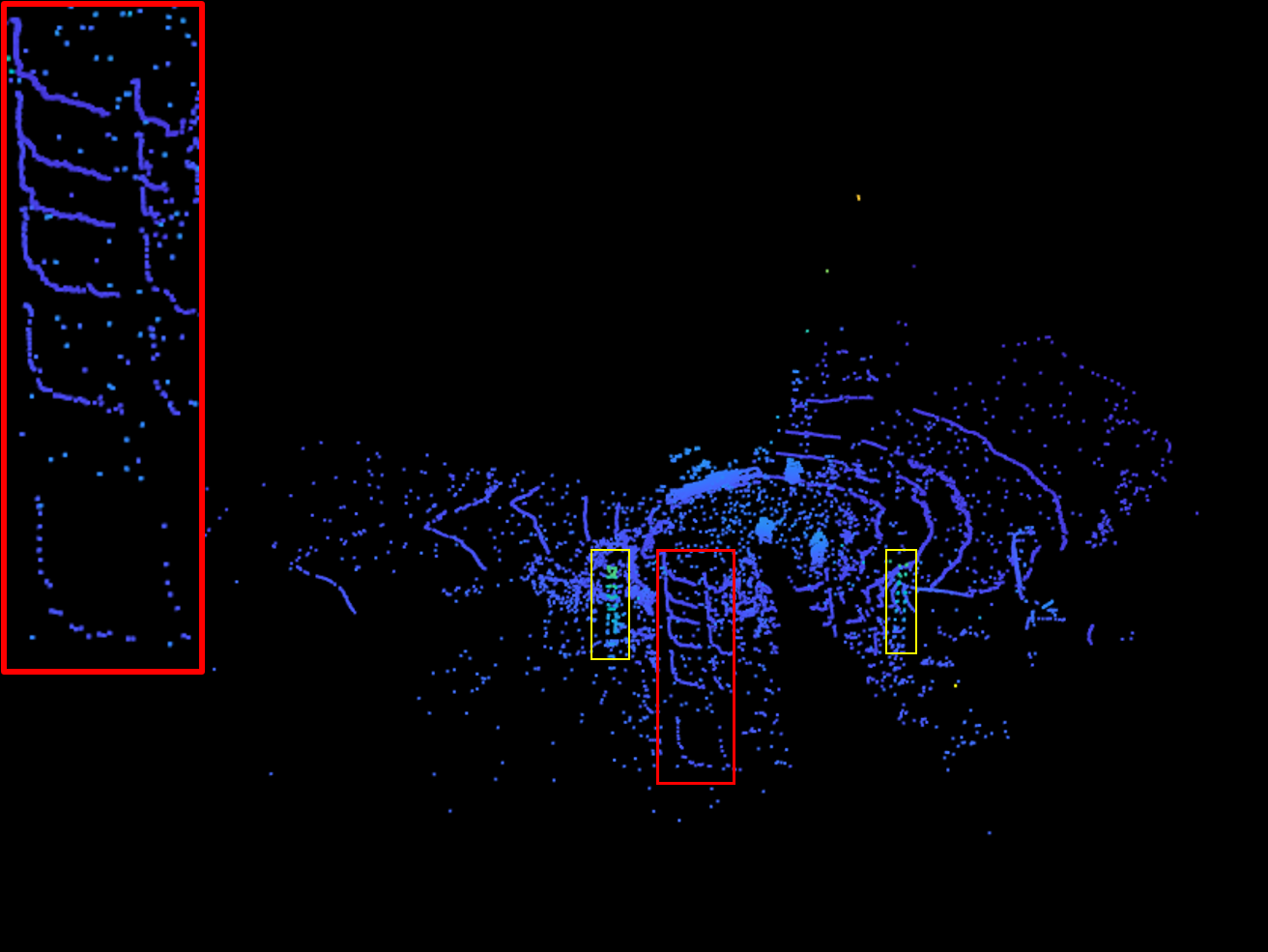}}\hfill
\subfloat[DSOR]{\includegraphics[width=0.32\textwidth]{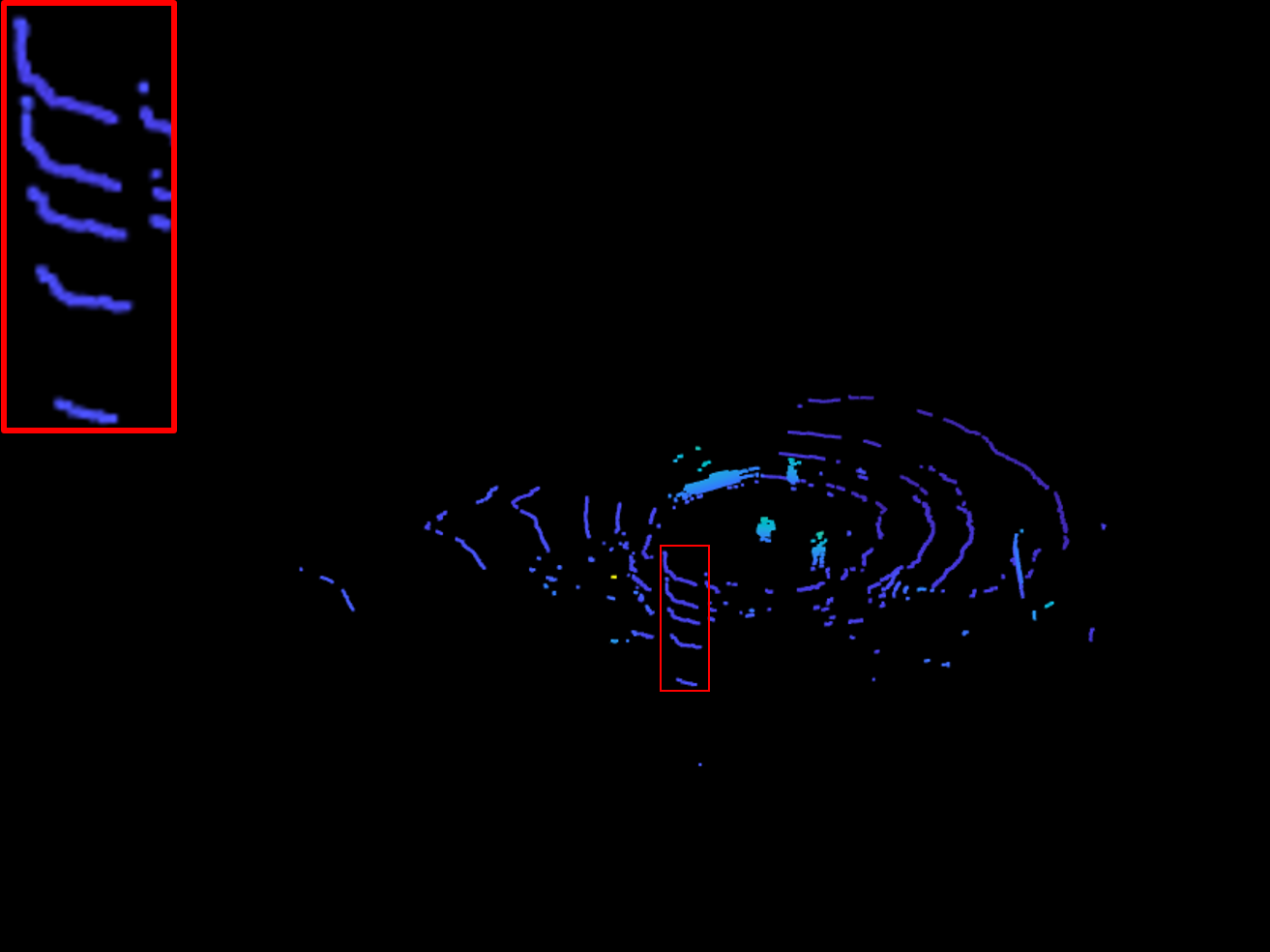}}\hfill
\subfloat[DROR]{\includegraphics[width=0.32\textwidth]{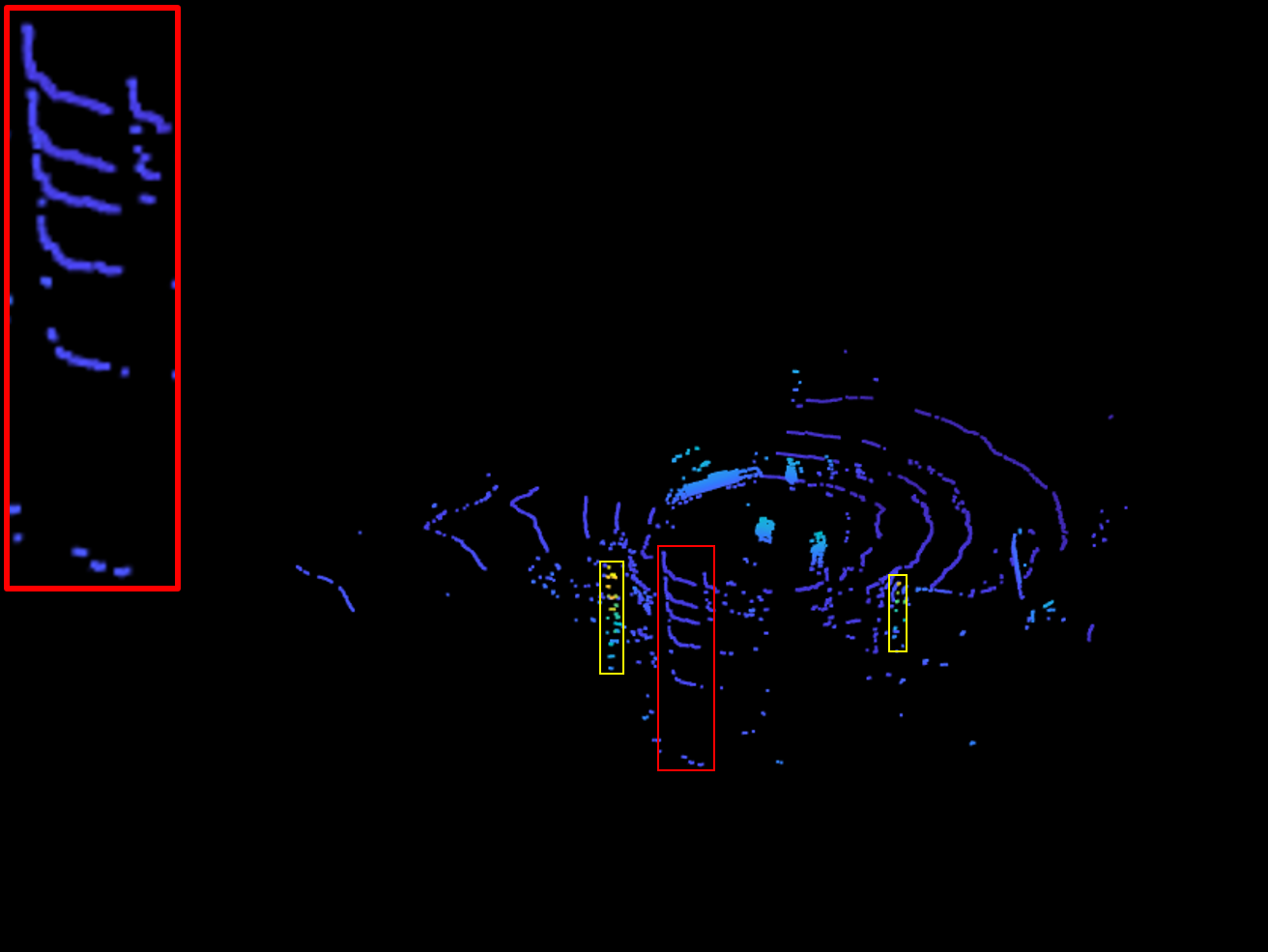}}
\caption{LiDAR point clouds at an experimental railway level crossing in Sweden. (a) Point cloud augmented with simulated rain at $50~\mathrm{mm/h}$ using~\cite{teufel2022simulating}. (b) DSOR result. (c) DROR result. Both methods suppress rain-induced reflections and preserve salient objects (vehicle and pedestrians), but degrade fine track details (red boxes; enlarged views in the upper-right corners), particularly DSOR; signal towers (yellow boxes) are removed by DSOR and partially degraded by DROR.}

\label{fig:LC_ptcloud}
\end{figure*}

Fig.~\ref{fig:LC_ptcloud}(a) shows a LiDAR point cloud collected using a Velodyne VLP-16 sensor at an experimental railway level crossing in Sweden and augmented with simulated rain at a rate of 50~mm/h using the method in~\cite{teufel2022simulating}. The corresponding denoised point clouds obtained using DSOR and DROR are shown in Fig.~\ref{fig:LC_ptcloud}(b) and (c), respectively. The scene includes a vehicle passing through the level crossing together with three pedestrians in the vicinity. Both DSOR and DROR suppress rain-induced reflections while preserving these salient objects. However, the railway tracks---highlighted by the red bounding boxes and shown in enlarged views at the upper-right corner---exhibit noticeable loss of fine structural details after denoising, particularly when DSOR is applied. Similarly, the signal towers on both sides of the railway, marked by the yellow bounding boxes, are completely removed by DSOR and partially degraded by DROR. In railway safety--critical applications, accurate preservation of track structures is essential, which motivates the development of a point cloud denoising algorithm that effectively suppresses rain-induced patterns while retaining fine structural details, without requiring cumbersome manual parameter tuning as in methods such as DDIOR.

The contributions of this paper can be summarize as:
\begin{itemize}
\item We introduce an empirical range-dependent distribution that characterizes LiDAR reflections induced by weather particles.

\item Based on this distribution, we propose an Intensity- and Distance-Aware Statistical Outlier Removal (IDSOR) method for weather-induced point cloud denoising. In addition, we develop a variant of IDSOR that does not rely on a predefined empirical distribution, but instead estimates the weather-particle distribution from an initial DROR-based filtering stage and subsequently applies a second IDSOR refinement.

\item We demonstrate that the proposed IDSOR methods outperform DSOR, DROR and DDIOR in terms of both recall and precision on real measured data, simulation-augmented data, as well as the Winter Adverse Driving Dataset (WADS)~\cite{kurup2021wads}.
\end{itemize}

\section{IDSOR Filter}
\subsection{LiDAR Returns in Adverse Weather}
\label{sec:2}

Returns induced by rain and snow particles exhibit a characteristic ``near-is-dense, far-is-sparse'' pattern, as reported in~\cite{wang2022scalable,gupta2025evaluating}. This observation suggests that a self-adaptive threshold that jointly exploits point density and intensity information can benefit from an accurate statistical description of the probability density function (PDF) of weather-particle-induced LiDAR returns.

\begin{figure}[t]
    \centering
    \includegraphics[width=\linewidth]{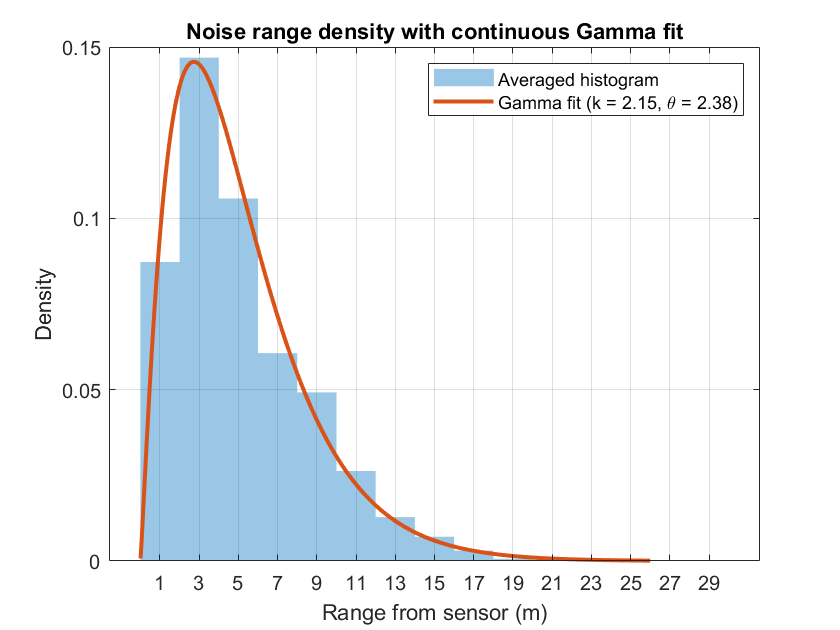}
    \caption{Averaged histograms of weather-particle-induced LiDAR returns at different ranges under simulated rain and snow conditions, and the corresponding Gamma fit.}
    \label{fig:pdf_gamma}
\end{figure}

To obtain such a description, we generate weather-contaminated point clouds using the model-based rain/snow augmentation procedure in \cite{teufel2022simulating} for multiple rain and snow rates. For each augmented scan, we extract the weather-induced points and compute their ranges to the sensor. We then aggregate the range samples across all simulated rates and form an averaged histogram over distance bins of width 3~m, as shown in Fig.~\ref{fig:pdf_gamma}. Since the range is non-negative and the empirical histograms exhibit a Gamma-like shape, we model the range distribution using a Gamma distribution,
\begin{equation}
f_r(r;k,\theta)
= \frac{1}{\Gamma(k)\,\theta^{k}}\, r^{k-1}
\exp\!\left(-\frac{r}{\theta}\right),\quad r\ge 0,
\end{equation}
where $r$ denotes the range from the LiDAR sensor, and $k$ and $\theta$ are the shape and scale parameters, respectively. In this study, the estimated parameters are $k=2.15$ and $\theta=2.38$.

\subsection{IDSOR Threshold}
IDSOR extends the DSOR filter~\cite{kurup2021dsor} by adaptively modulating the decision threshold using both intensity cues and spatial sparsity. Intuitively, points with higher intensity are more likely to correspond to true object returns and should therefore be assigned a larger threshold to reduce the risk of being erroneously removed. Moreover, the relative contribution of intensity versus geometry is made range-dependent. In the near range, weather-induced returns are typically dense around the sensor and exhibit relatively low intensities~\cite{gupta2025evaluating,teufel2022simulating}, making intensity particularly discriminative; hence, IDSOR places stronger emphasis on intensity. In the far range, weather-induced outliers become sparse, while the intensities of true returns may also decrease due to range-dependent attenuation and reflectance effects. Consequently, IDSOR progressively shifts the decision rule toward geometric consistency (i.e., neighborhood sparsity) to avoid over-penalizing low-intensity true returns at longer ranges.

The IDSOR threshold is constructed as follows. First, we map the range-dependent outlier PDF to a logistic-like weight in $[0,1]$:
\begin{equation}
\alpha_i = \frac{\rho\, f_r(r_i)}{\rho\, f_r(r_i) + 1},
\end{equation}
where $r_i$ denotes the range of the $i$th point, $f_r(\cdot)$ is the fitted range-dependent PDF of weather-induced returns, and $\rho$ is a tunable parameter capturing the weather severity (e.g., rain rate). This monotonic mapping satisfies $\alpha_i \to 1$ as $\rho f_r(r_i)\to \infty$ and $\alpha_i \to 0$ as $\rho f_r(r_i)\to 0$. Hence, larger $\rho$ or $f_r(r_i)$ yields a larger $\alpha_i$, indicating that the thresholding rule should place more emphasis on intensity at range $r_i$.

Next, the LiDAR intensity values are normalized to $[0,1]$ by dividing by the maximum intensity in the scan, i.e., $i_{\mathrm{norm},i}\in[0,1]$. We then define an intensity-based score
\begin{equation}
h_i = 1 - i_{\mathrm{norm},i},
\end{equation}
so that a larger $h_i$ corresponds to a lower measured intensity and, hence, a higher likelihood of being a weather-induced return.

Finally, the IDSOR threshold for the $i$th point is defined as
\begin{equation}
T_{\mathrm{IDSOR}} = s\,T_g \bigl(1-\alpha_i h_i\bigr),
\label{threshold}
\end{equation}
where $T_g$ is the global SOR threshold~\cite{rusu20113d} and $s$ is a multiplicative scaling factor. A larger $s$ increases $T_{\mathrm{IDSOR},i}$ and yields a less aggressive filter, whereas a smaller $s$ decreases $T_{\mathrm{IDSOR},i}$ and results in more aggressive filtering. From \eqref{threshold}, higher intensity (larger $i_{\mathrm{norm},i}$, hence smaller $h_i$) increases $T_{\mathrm{IDSOR},i}$, making strong returns less likely to be removed, while lower intensity decreases $T_{\mathrm{IDSOR},i}$ and promotes rejection of weak returns. The range effect is governed by $\alpha_i$: at ranges where weather-induced returns are expected to be dense (larger $f_r(r_i)$ or $\rho$), $\alpha_i$ increases and the threshold becomes more intensity-driven; at farther ranges, $\alpha_i$ decreases and $T_{\mathrm{IDSOR},i}$ approaches the baseline $sT_g$, thereby relying more on geometric consistency.

\begin{algorithm}[t]
\caption{Intensity- and Distance-Aware Statistical
Outlier Removal (IDSOR) Filter (using the notation of~\cite{kurup2021dsor})}
\label{alg:idsor}
\begin{algorithmic}[1]
\REQUIRE Point cloud $\mathbf{P}=\{\mathrm{p}_i\}_{i=1}^{N}$, $\mathrm{p}_i=[x_i,y_i,z_i,i_{\text{norm},i}]^T$; \\
Empirical outlier range-dependent PDF $f_r(r;k,\theta)$ \\
        $k_\text{min}$: minimum number of nearest neighbors; \\
        $s_g$:  multiplication factor for standard deviation
factor; \\
$\rho$: tunable parameter capturing the weather severity; \\
$s$: multiplication factor for the threshold. 

\ENSURE Filtered point cloud $\mathbf{F} = \{f_i\}_{i=1}^{N}$; $\mathrm{f}_i=[x_i,y_i,z_i]^T$.

\STATE $P\leftarrow KdTree$
\FOR{$p_i \in P$} 
    \STATE $mean\_distances \ (\bar d_i)\leftarrow nearestKSearch(k_\text{min})$
\ENDFOR

\STATE mean $\mu \leftarrow \frac{1}{N}\sum_{i=1}^{N}\bar d_i$
\STATE standard deviation $\sigma \leftarrow \sqrt{\frac{1}{N}\sum_{i=1}^{N}(\bar d_i-\mu)^2}$
\STATE global threshold $T_g \leftarrow \mu + s_g\,\sigma$

\FOR{$p_i \in P$}
    \STATE $r_i \leftarrow \sqrt{x_i^2+y_i^2+z_i^2}$
    \STATE $\alpha_i = \frac{\rho\, f_r(r_i)}{\rho\, f_r(r_i) + 1}$
    \STATE $h_i = 1 - i_{\mathrm{norm},i}$
    \STATE $T_{\mathrm{IDSOR}} \leftarrow T_g s  \bigl(1-\alpha_i h_i\bigr)$
    \IF{$\bar d_i < T_{\mathrm{IDSOR}}$}
        \STATE $f_i \leftarrow p_i  \qquad\{Inliers \}$  
    \ENDIF
\ENDFOR

\RETURN $\mathbf{F}$
\end{algorithmic}
\end{algorithm}

\subsection{A DROR-Prior Variant of IDSOR}
We also consider a DROR-prior variant of IDSOR, where DROR is used only to facilitate the estimation of the weather-return distribution. Specifically, DROR \cite{charron2018noising} is applied as a coarse selector to identify a subset of points that are likely dominated by weather-induced returns. These selected points are then used to estimate a range-dependent PDF, denoted by $\hat f_r(\cdot)$, which serves as a data-driven prior in the construction of $\alpha_i$ and the IDSOR threshold. Finally, IDSOR is applied to the original point cloud using $\hat f_r(\cdot)$ in place of the empirical $f_r(\cdot)$.

\section{EVALUATION AND RESULTS}
We evaluate DSOR, DROR, DDIOR, the proposed IDSOR, and its DROR-prior variant on simulation-augmented data collected at a railway level crossing, as well as on the Winter Adverse Driving Dataset (WADS)~\cite{kurup2021wads}. Each filters was individually tuned to remove most snow returns while preserving as much of the environment as possible, except DDIOR. Since DDIOR requires distance-wise manual tuning of weighting parameters, which incurs a high calibration cost, we directly adopt the parameter settings recommended in~\cite{wang2022scalable}. We report both qualitative and quantitative comparisons. The qualitative evaluation is based on visual inspection, with particular emphasis on preserving safety-critical track structures in the level-crossing scenario. For quantitative evaluation, we report precision and recall.

\subsection{Qualitative evaluation}
\begin{figure*}
\centering
\subfloat[DDIOR]{\includegraphics[width=0.32\textwidth]{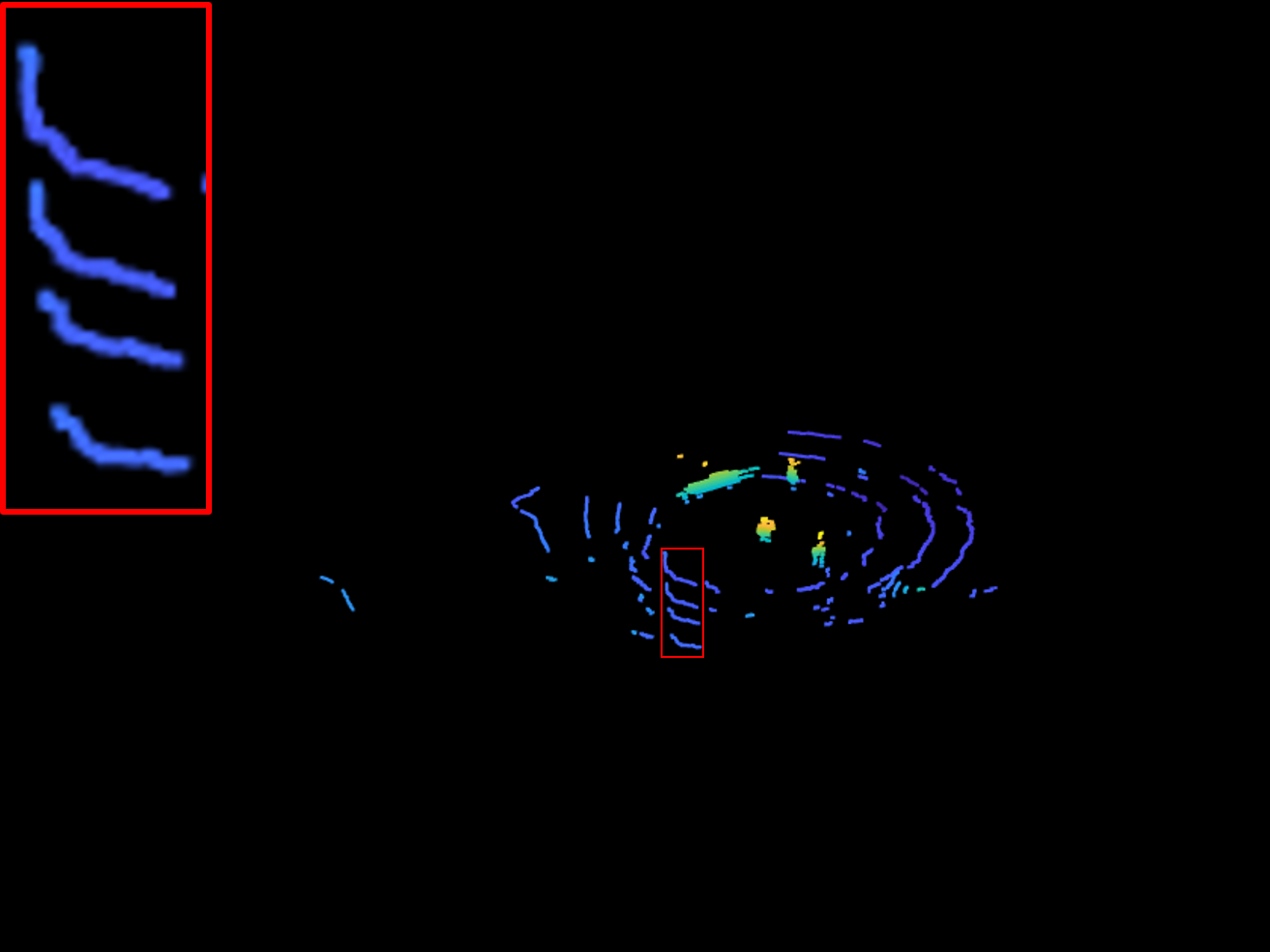}}\hfill
\subfloat[IDSOR]{\includegraphics[width=0.32\textwidth]{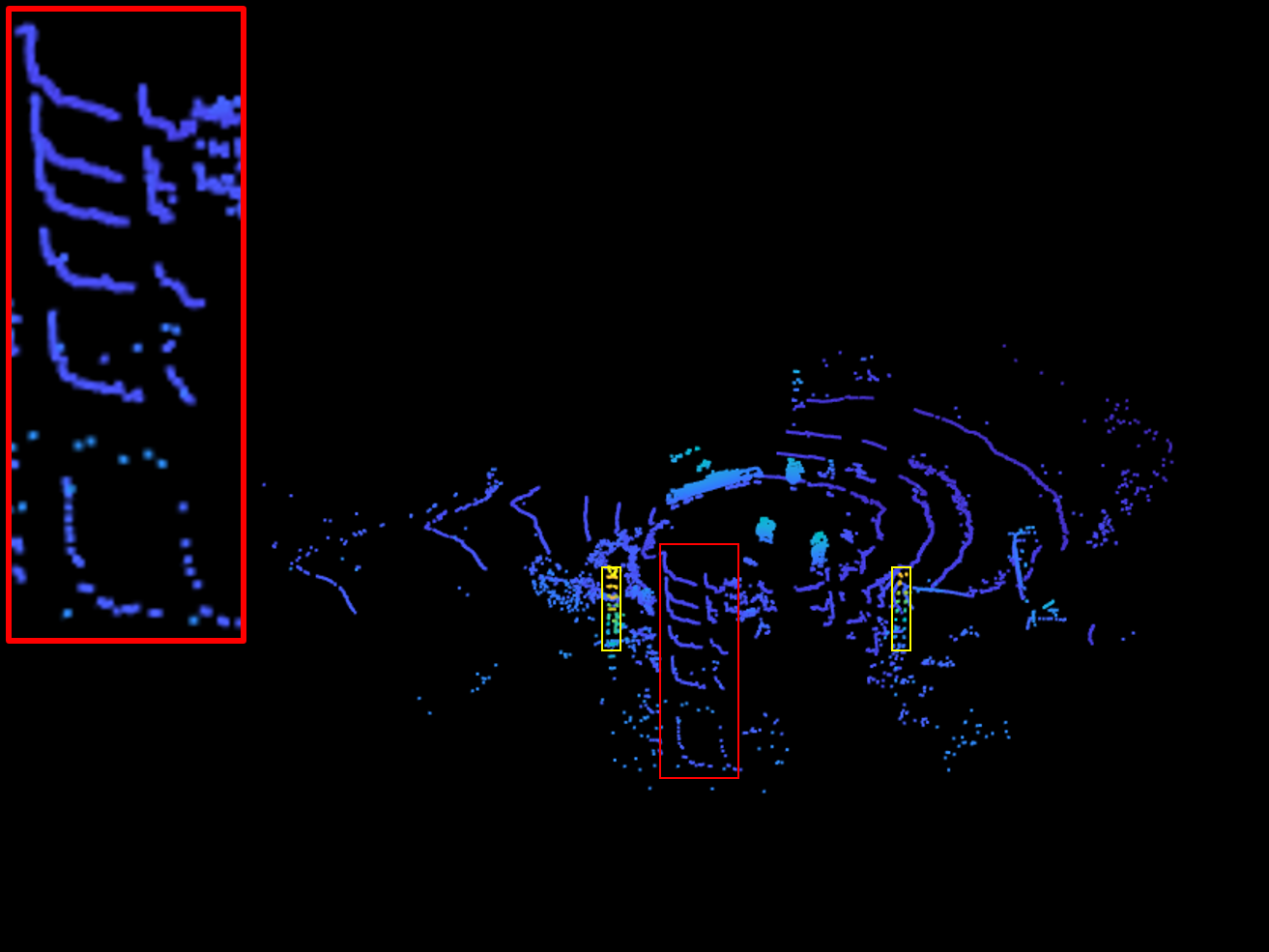}} \hfill
\subfloat[DROR-prior IDSOR]{\includegraphics[width=0.32\textwidth]{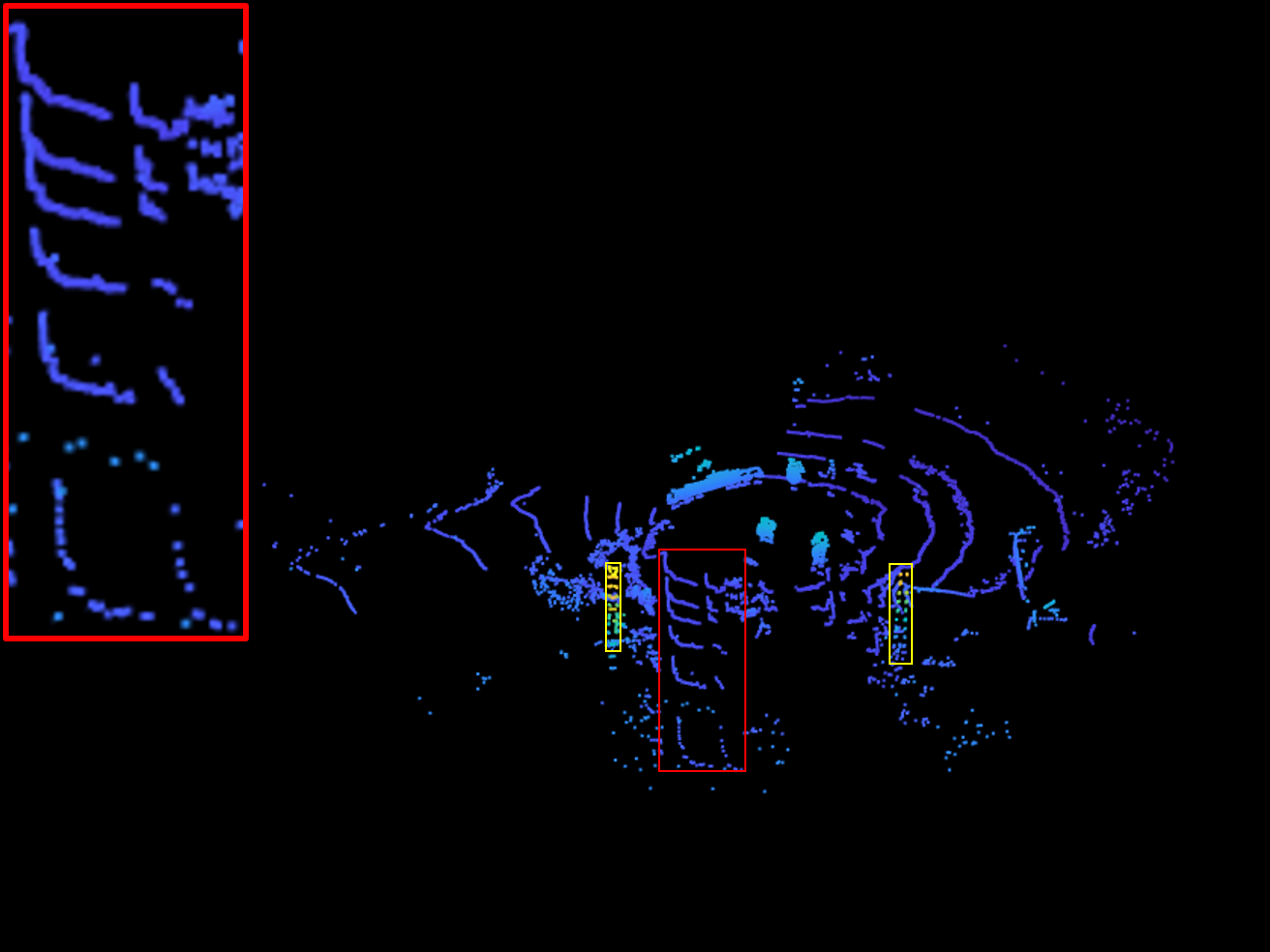}}\hfill
\caption{Level-crossing dataset under simulated rain (50~mm/h): (a)~DDIOR, (b)~IDSOR, and (c)~DROR-prior IDSOR. Red bounding boxes (with enlarged views in the upper-right corners) indicate the railway tracks, and yellow bounding boxes indicate the signal towers.}
\label{fig:LC_cloud_filtered}
\end{figure*}

As shown in Fig.~\ref{fig:LC_cloud_filtered}, for the level-crossing data, both DSOR and DROR degrade fine structural details of the railway tracks; moreover, the signal towers are completely removed by DSOR and partially degraded by DROR. We further compare DDIOR, the proposed IDSOR, and its DROR-prior variant in Fig.~\ref{fig:LC_cloud_filtered}. It can be observed that IDSOR and the DROR-prior IDSOR suppress most weather-induced returns in the vicinity of the level crossing while largely preserving the railway tracks and the signal towers. In contrast, under the parameter settings recommended in~\cite{wang2022scalable}, DDIOR exhibits a noticeable loss of these structures.

In addition to our level-crossing measurements, we also evaluate the proposed method on the Winter Adverse Driving Dataset (WADS) \cite{kurup2021wads}. WADS follows the KITTI-style point-wise labeling protocol~\cite{behley2019iccv,geiger2012cvpr} and additionally includes two weather-specific classes: accumulated snow and falling snow. In our experiments, we select sequences containing only falling snow, as the current method is designed for suppressing airborne particle returns and is not yet tailored to removing accumulated snow. The same sequence, comprising 170{,}882 labeled points, is used for both qualitative visualization and the quantitative evaluation.

\begin{figure*}[!t]
\centering
\subfloat[Original point cloud]{\includegraphics[width=0.32\textwidth]{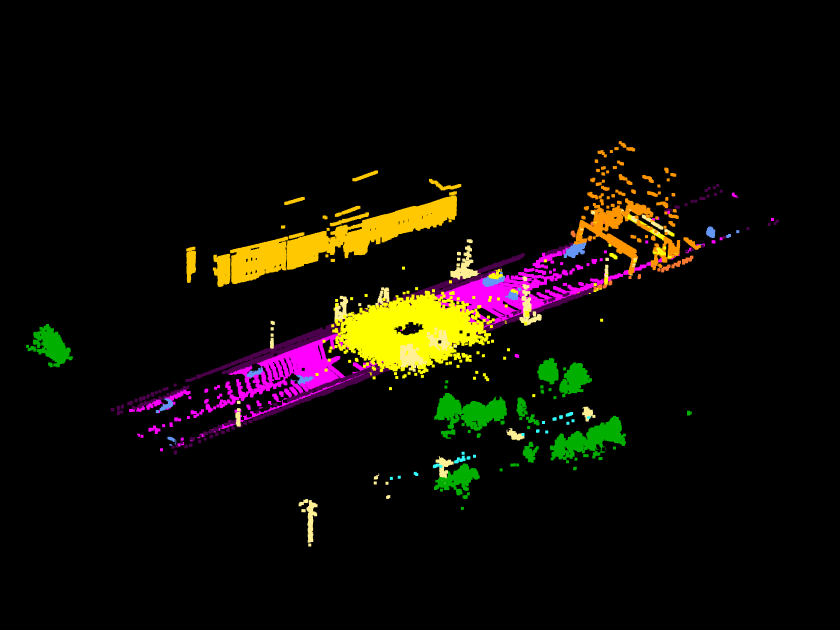}}\hfill
\subfloat[DSOR]{\includegraphics[width=0.32\textwidth]{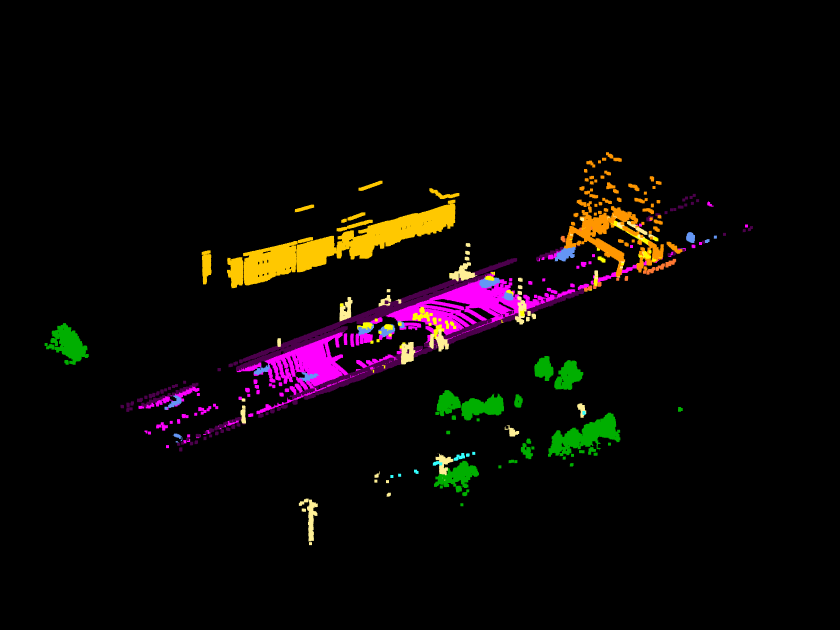}} \hfill
\subfloat[DROR]{\includegraphics[width=0.32\textwidth]{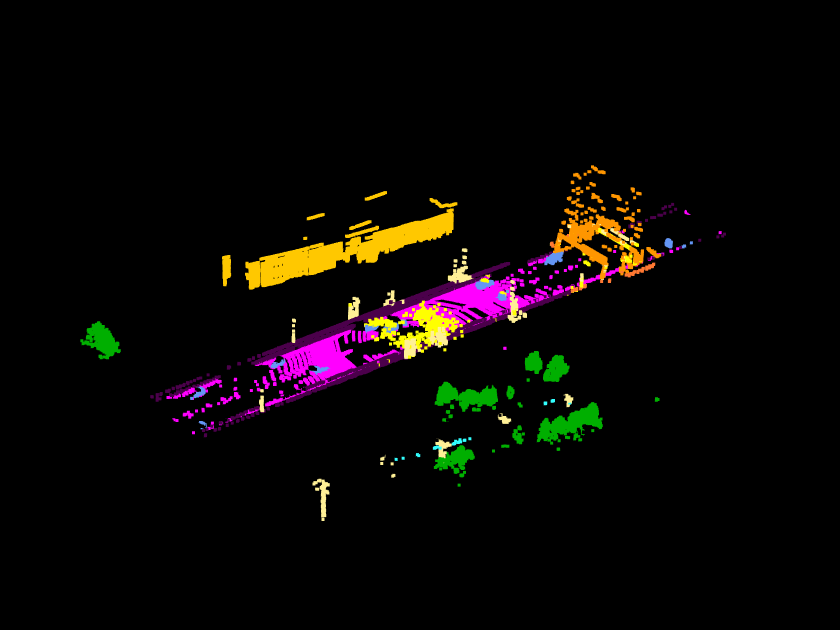}} \\
\subfloat[DDIOR]{\includegraphics[width=0.32\textwidth]{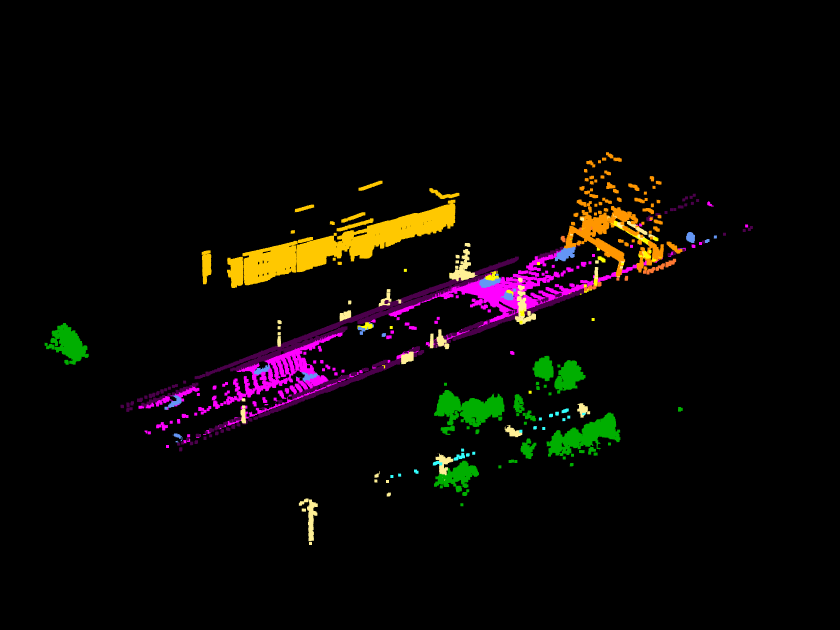}} \hfill
\subfloat[IDSOR]{\includegraphics[width=0.32\textwidth]{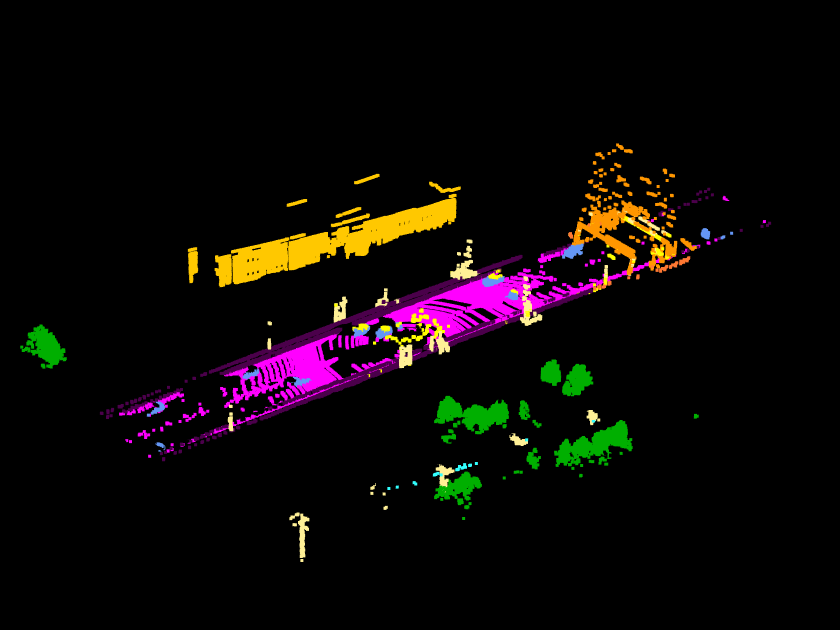}} \hfill
\subfloat[DROR-prior IDSOR]{\includegraphics[width=0.32\textwidth]{graph/IDSOR_DROR.png}}
\caption{Qualitative comparison on WADS~\cite{kurup2021wads}. The yellow points highlight the falling-snow class in the WADS annotations. From left to right and top to bottom: (a) original, (b) DSOR, (c) DROR, (d) DDIOR, (e) IDSOR, and (f) DROR-prior IDSOR.}
\label{fig:WAD}
\end{figure*}

Fig.~\ref{fig:WAD} shows the original point cloud and the filtered results obtained using DSOR, DROR, DDIOR, IDSOR, and the DROR-prior IDSOR variant. The yellow points concentrated around the center of the road correspond to the falling-snow class in the WADS annotations. Visually, IDSOR and DSOR exhibit comparable outlier-suppression performance, whereas DROR leaves noticeably more residual outliers. In contrast, DDIOR removes weather-induced points more aggressively, but at the expense of attenuating or removing a substantial number of returns from physical objects.

\begin{figure}
    \centering
    \includegraphics[width=\linewidth]{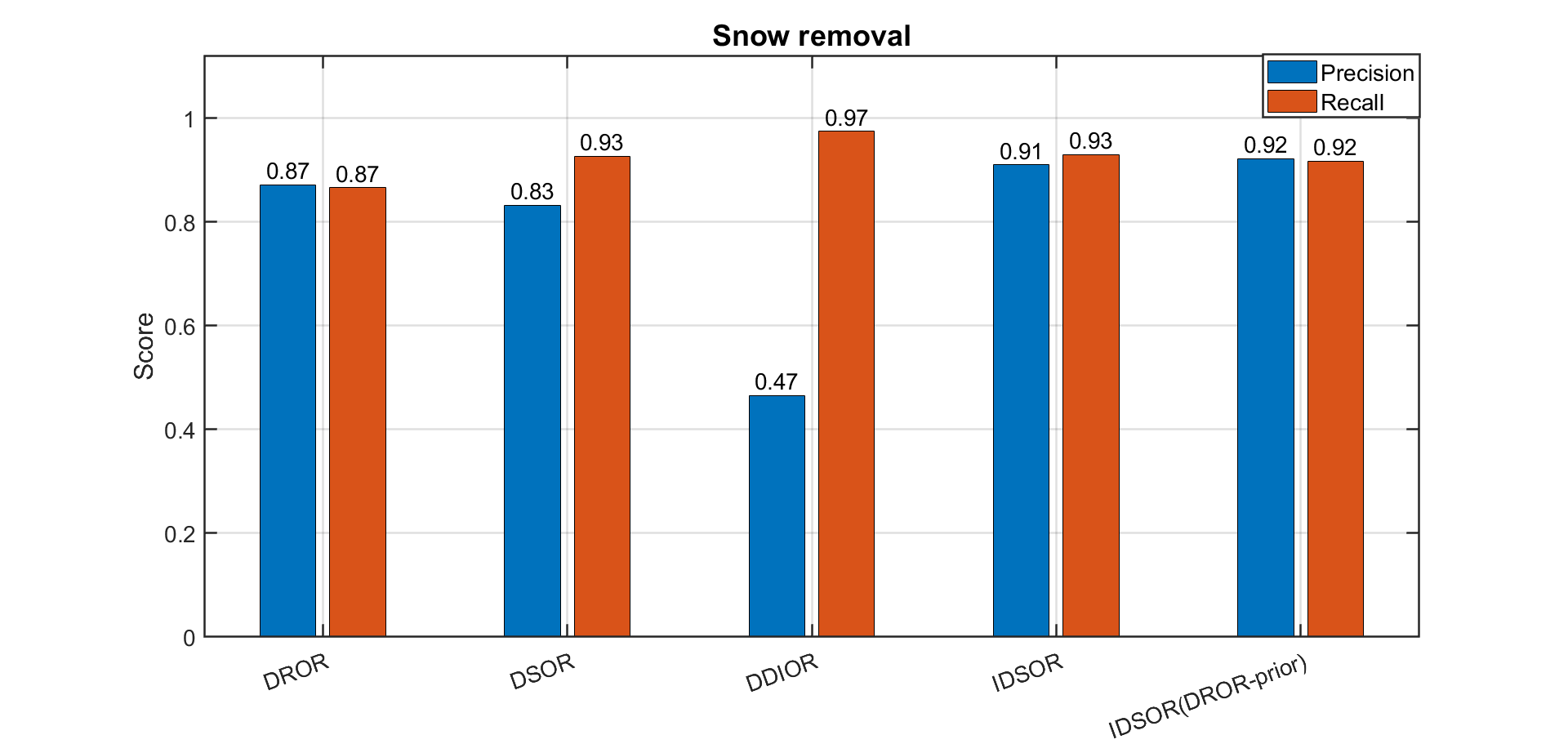}
    \caption{Precision and recall comparison of DROR, DSOR, DDIOR, IDSOR, and DROR-prior IDSOR for falling-snow removal on the WADS dataset.}

    \label{fig:bar}
\end{figure}

\subsection{Quantitative evaluation}
To quantitatively assess denoising performance, we report precision and recall, which characterize the trade-off between suppressing weather-induced points and preserving true scene returns. We treat weather-induced points (e.g., rain/snow returns) as the positive class to be removed. Recall measures the fraction of ground-truth weather points that are successfully removed (higher recall implies fewer residual outliers), whereas precision measures the fraction of removed points that truly correspond to weather returns (higher precision implies fewer true scene points being mistakenly removed).

Fig.~\ref{fig:bar} compares precision and recall of DROR, DSOR, DDIOR, IDSOR, and DROR-prior IDSOR for snow removal. DROR is moderate (0.87/0.87), while DSOR increases recall (0.93) at reduced precision (0.83). DDIOR yields the highest recall (0.97) but very low precision (0.47), indicating excessive removal of object returns. IDSOR achieves the best balance with high precision and recall (0.91/0.93). The DROR-prior variant is slightly more conservative, improving precision (0.92) while lowering recall (0.92).

\section{Conclusion}
We proposed IDSOR, an intensity- and distance-aware LiDAR denoising method that exploits the range-dependent distribution of weather-induced returns. By incorporating the range distribution into an adaptive threshold, IDSOR avoids the manual tuning required by intensity-augmented baselines such as DDIOR and achieves consistently improved performance on both the level-crossing and WADS datasets, with over $90\%$ precision and recall on WADS.

\bibliographystyle{IEEEtran}
\bibliography{ref}

@article{li2020lidar,
  title={Lidar for autonomous driving: The principles, challenges, and trends for automotive lidar and perception systems},
  author={Li, You and Ibanez-Guzman, Javier},
  journal={IEEE Signal Processing Magazine},
  volume={37},
  number={4},
  pages={50--61},
  year={2020},
  publisher={IEEE}
}

@article{mulder2011use,
  title={The use of remote sensing in soil and terrain mapping—A review},
  author={Mulder, VL and De Bruin, S and Schaepman, Michael E and Mayr, TR},
  journal={Geoderma},
  volume={162},
  number={1-2},
  pages={1--19},
  year={2011},
  publisher={Elsevier}
}

@inproceedings{gargoum2017automated,
  title={Automated extraction of road features using LiDAR data: A review of LiDAR applications in transportation},
  author={Gargoum, Suliman and El-Basyouny, Karim},
  booktitle={2017 4th International Conference on Transportation Information and Safety (ICTIS)},
  pages={563--574},
  year={2017},
  organization={IEEE}
}

@article{yue2024lidar,
  title={LiDAR-based {SLAM} for robotic mapping: state of the art and new frontiers},
  author={Yue, Xiangdi and Zhang, Yihuan and Chen, Jiawei and Chen, Junxin and Zhou, Xuanyi and He, Miaolei},
  journal={Industrial Robot: the international journal of robotics research and application},
  volume={51},
  number={2},
  pages={196--205},
  year={2024},
  publisher={Emerald Publishing Limited}
}

@inproceedings{charron2018noising,
  title={De-noising of lidar point clouds corrupted by snowfall},
  author={Charron, Nicholas and Phillips, Stephen and Waslander, Steven L},
  booktitle={2018 15th Conference on Computer and Robot Vision (CRV)},
  pages={254--261},
  year={2018},
  organization={IEEE}
}

@inproceedings{rusu20113d,
  title={3{D} is here: Point cloud library {(PCL)}},
  author={Rusu, Radu Bogdan and Cousins, Steve},
  booktitle={2011 IEEE international conference on robotics and automation},
  pages={1--4},
  year={2011},
  organization={IEEE}
}

@article{balta2018fast,
  title={Fast statistical outlier removal based method for large 3{D} point clouds of outdoor environments},
  author={Balta, Haris and Velagic, Jasmin and Bosschaerts, Walter and De Cubber, Geert and Siciliano, Bruno},
  journal={IFAC-PapersOnLine},
  volume={51},
  number={22},
  pages={348--353},
  year={2018},
  publisher={Elsevier}
}

@article{kurup2021dsor,
  title={{DSOR}: A scalable statistical filter for removing falling snow from lidar point clouds in severe winter weather},
  author={Kurup, Akhil and Bos, Jeremy},
  journal={arXiv preprint arXiv:2109.07078},
  year={2021}
}

@inproceedings{teufel2022simulating,
  title={Simulating realistic rain, snow, and fog variations for comprehensive performance characterization of lidar perception},
  author={Teufel, Sven and Volk, Georg and Von Bernuth, Alexander and Bringmann, Oliver},
  booktitle={2022 IEEE 95th Vehicular Technology Conference:(VTC2022-Spring)},
  pages={1--7},
  year={2022},
  organization={IEEE}
}

@article{wang2022scalable,
  title={A scalable and accurate de-snowing algorithm for LiDAR point clouds in winter},
  author={Wang, Weiqi and You, Xiong and Chen, Lingyu and Tian, Jiangpeng and Tang, Fen and Zhang, Lantian},
  journal={Remote Sensing},
  volume={14},
  number={6},
  pages={1468},
  year={2022},
  publisher={MDPI}
}

@article{gupta2025evaluating,
  title={Evaluating LiDAR Perception Algorithms for All-Weather Autonomy},
  author={Gupta, Himanshu and Lilienthal, Achim J and Andreasson, Henrik},
  journal={Sensors},
  volume={25},
  number={24},
  pages={7436},
  year={2025},
  publisher={MDPI}
}

@misc{kurup2021wads,
  author       = {Kurup, Aravind and Bos, Jennifer},
  title        = {{The Winter Adverse Driving Dataset (WADS)}},
  year         = {2021},
  howpublished = {Dataset, Michigan Technological University},
  url          = {https://digitalcommons.mtu.edu/wads/},
  urldate      = {2026-01-13}
}

@inproceedings{behley2019iccv,
  author = {J. Behley and M. Garbade and A. Milioto and J. Quenzel and S. Behnke and C. Stachniss and J. Gall},
  title = {{SemanticKITTI: A Dataset for Semantic Scene Understanding of LiDAR Sequences}},
  booktitle = {Proc. of the IEEE/CVF International Conf.~on Computer Vision (ICCV)},
  year = {2019}
}

@inproceedings{geiger2012cvpr,
  author = {A. Geiger and P. Lenz and R. Urtasun},
  title = {{Are we ready for Autonomous Driving? The KITTI Vision Benchmark Suite}},
  booktitle = {Proc.~of the IEEE Conf.~on Computer Vision and Pattern Recognition (CVPR)},
  pages = {3354--3361},
  year = {2012}
}

@misc{yan2026obstacledetectionlevelcrossings,
  author = {Chenyang Yan and Mats Bengtsson},
  title  = {Obstacle Detection at Level Crossings under Adverse Weather Conditions -- A Survey},
  year   = {2026},
  note   = {arXiv:2602.01974},
  url    = {https://arxiv.org/abs/2602.01974}
}

\end{document}